\documentclass[prl,twocolumn,tightenlines,superscriptaddress,a4paper,lengthcheck]{revtex4-1}
\usepackage{amsfonts}
\usepackage{amsmath}
\usepackage{amssymb}
\usepackage{graphicx}
\usepackage{graphicx,amssymb,amsbsy,color}
\usepackage{bm}
\usepackage{bbold}
\usepackage{color}
\providecommand{\U}[1]{\protect\rule{.1in}{.1in}}

\begin{document}

\title{Interband Theory of Kerr Rotation in Unconventional Superconductors}
\author{Robert Joynt}
\affiliation{Department of Physics, University of Wisconsin-Madison, Madison, WI 53706, USA}
\affiliation{International Center for Quantum Materials, School of Physics, Peking University, Beijing 100871, China}
\author{Wen-Chin Wu}
\affiliation{Department of Physics, National Taiwan Normal University, Taipei 11677, Taiwan}
\date{\today }

\begin{abstract}
Recent experiments have shown rotation of the plane of polarization of light
reflected from the surface of some superconductors. \ This indicates that
time reversal and certain mirror symmetries are broken in the ordered phase.
\ The photon energy exceeds the electronic bandwidth, so that completely
filled or completely empty bands must play a role. \ We show that in
strong-coupling theory a Coulomb interaction may produce an order parameter
in the unoccupied band that explains the observations. \ The theory puts
tight constraints on the form of the order parameter in different bands. \
We propose that the experiments have detected, for the first time, \ the
existence of a superconducting order parameter in a band far from the Fermi
energy.
\end{abstract}

\maketitle

The subject of unconventional superconductivity is now over 30 years old,
and the prime driving force in the field has been the determination of the
form of the order parameter. \ This issue has become more urgent in the era
of topological superconductivity. \ The determination has generally proved
to be surprisingly difficult: unambiguous identifications remain remarkably
few in number. \ For this reason, it is useful to have experiments that test
symmetry breaking directly, rather than after a train of reasoning and
assumptions. \ For time-reversal symmetry breaking, conceptually the
simplest such experiment is Kerr rotation \cite{1367-2630-11-5-055060}, the
rotation of the plane of polarization of reflected light through an angle $%
\theta _{K}$. \ This effect in superconductors was already active in
experiments \cite{PhysRevLett.65.123} and theory \cite%
{PhysRevB.44.4720,Yip1992} in the 1990s. \ \ However, positive experimental
results are relatively recent. \ A nonzero signal that begins at the onset
of superconductivity and grows as temperature decreases has been observed in
Sr$_{2}$RuO$_{4}$ \cite{PhysRevLett.97.167002}, UPt$_{3}$ \cite{Schemm190},
and URu$_{2}$Si$_{2}$ \cite{PhysRevB.91.140506}. The simplest theories of
the pure system do not give a result large enough to explain the observed
magnitude of $\theta _{K}\approx 10^{-6}$. \ In\ Sr$_{2}$RuO$_{4}$ this has
been interpreted in different ways. \ One is to invoke impurity scattering
\cite{PhysRevB.78.060501,PhysRevB.80.104508}. \ Another is to attribute the
effect to interband transitions between bands that cross the Fermi energy
\cite{PhysRevLett.108.077004,PhysRevLett.108.157001}. This paper focuses on
clean UPt$_{3},$ though we comment on other compounds below. \ Since in UPt$%
_{3}$ the bandwidth is less than the energy (0.8 eV) of the light \cite%
{Oguchi1985,PhysRevB.35.7260}, the transitions induced by the light can only
be of an interband nature. \ UPt$_{3}$ is the most clear-cut example of a
material in which purely intraband effects cannot account for the
observations. \ We propose that the Coulomb interaction induces a nonzero
order parameter in the completely unoccupied and completely occupied bands
that is responsible for the Kerr effect. \ This implies that the Kerr effect
experiments probe, for the first time, a superconducting order parameter in
a completely unoccupied or completely occupied band distant from the Fermi
energy. \ A somewhat similar effect was proposed for semiconductors with
small gaps \cite{Nozieres1985} and may have been seen in a completely
occupied band relatively near the Fermi energy in LiFeAs \cite%
{PhysRevB.91.161108}.

Reflection of light of frequency $\omega$ incident on a sample along the $z$%
-axis is controlled by the components $\varepsilon_{xx}\left( \omega\right)$%
, $\varepsilon_{yy}\left( \omega\right)$, $\varepsilon _{xy}\left(
\omega\right)$, and $\varepsilon_{yx}\left( \omega\right)$ of the dielectric
tensor. \ UPt$_{3}$ is hexagonal and we can assume that tetragonal
distortions that give rise to the inequality of $\varepsilon _{xx}\left(
\omega\right) $ and $\varepsilon_{yy}\left( \omega\right) $ (linear
birefringence) are absent so $\varepsilon_{xx}\left( \omega\right)
=\varepsilon_{yy}\left( \omega\right) =\varepsilon$. \ In any case, the
experiments are designed not to be sensitive to linear birefringence \cite%
{1367-2630-11-5-055060}. \ In a field $\vec{u}$ that breaks time reversal
symmetry, including an effective field from an order parameter, the Onsager
relation is $\varepsilon_{xy}\left( \omega,\vec{u}\right) $ $=\varepsilon
_{yx}\left( \omega,-\vec{u}\right) $ \cite{ecm} and since the assumption of
a linear relation between $\varepsilon $ and $\vec{u}$ is well-founded in
the present case, we find $\varepsilon_{xy}\left( \omega\right) $ $%
=-\varepsilon _{yx}\left( \omega\right)$. \ Then the normal modes in the
metal for a propagation direction $\vec{k}=k\hat{z}$ are circularly
polarized. \ For each frequency there are two wavevectors $k_{+}$ and $k_{-}$
that correspond to the two helicities: $k_{\pm}^{2}=\left(
\omega^{2}/c^{2}\right) \varepsilon _{\pm}$ and $\varepsilon_{\pm}\left(
\omega\right) =\varepsilon_{xx}\left( \omega\right) \pm i\left\vert
\varepsilon_{xy}\left( \omega\right) \right\vert$. The different dispersion
and absorption for $k_{+}$ and $k_{-}$ gives the Kerr rotation $\theta_{K}=%
\mathrm{Re}\left( \varepsilon _{xy}/\varepsilon_{xx}^{3/2}\right) $ in a
metal (for which $\left\vert \varepsilon_{xx}\right\vert \gg 1)$.\ To have $%
\varepsilon_{xy}\neq0$ or $\varepsilon_{yx}\neq0$, we need breaking of both
mirror symmetries $x\rightarrow-x$ and $y\rightarrow-y$ and time-reversal
\cite{PhysRevB.90.205130}. \ \ For a spatially uniform conventional singlet
superconductor with order parameter $\Delta$, time-reversal means $%
\Delta\rightarrow\Delta^{\ast}$ and a change of gauge is $%
\Delta\rightarrow\Delta e^{i\phi},$ so any time-reversal transformation is
equivalent to a gauge transformation and as a result there is no non-trivial
notion of time-reversal. \ For an unconventional superconductor with a
momentum-dependent order parameter, this is not the case: we might have $%
\Delta(\mathbf{p}) =c\Delta_{0}p_{z}\left( p_{x}+ip_{y}\right)$. \ Then $%
\Delta^{\ast}(\mathbf{p}) =c\Delta_{0}p_{z}\left( p_{x}-ip_{y}\right) $ and
no uniform phase factor relates $\Delta$ and $\Delta^{\ast}$.

An appropriate model Hamiltonian for the multiband superconductor UPt$_{3}$
is:%
\begin{widetext}
\begin{eqnarray}
H-\mu N &=& \sum_{n,\vec{p},\sigma}\xi\left(  n,\vec{p}\right)  a_{\sigma}^{\dag
}\left(  n,\vec{p}\right)  a_{\sigma}\left(  n,\vec{p}\right)\nonumber\\
&+&\sum_{n,\vec{p}}\sum_{n^{\prime},\vec{p}^{\prime}}
V\left(  n,\vec{p};n^{\prime},\vec{p}^{\prime}\right)
~a_{\uparrow}^{\dag}\left(  n,\vec{p}\right)
a_{\downarrow}^{\dag}\left(  n,-\vec{p}\right)  a_{\downarrow}\left(
n^{\prime},-\vec{p}^{\prime}\right)  a_{\uparrow}\left(  n^{\prime},\vec
{p}^{\prime}\right),
\label{eq:H}
\end{eqnarray}
\end{widetext}
where $\xi \left( n,\vec{p}\right) $\ are the single-particle energies
measured relative to the chemical potential, $\sigma $ is the pseudospin, $n$
and $n^{\prime }$ are band indices, $a_{\sigma }^{\dag }\left( n,\vec{p}%
\right) $ creates an electron in the state $n\vec{p}\sigma , $ and $V\left(
n,\vec{p};n^{\prime },\vec{p}^{\prime }\right) $ is a singlet pairing
interaction. \ The sum runs over all bands within 0.8 eV of the Fermi
energy. \ Since the bandwidths are of order $B\approx $ 0.2 to 0.3 eV, this
includes completely full and completely empty bands as well as the usual
partially occupied bands. \ Because of the narrow bandwidth, there is no
pair of partially occupied bands that have energies as much as 0.8 eV apart.
\

We have restricted our model to give singlet pairing only for ease of
presentation. \ The conclusions are essentially the same for triplet
pairing. $\ H$ is treated in the mean field approximation in a
straightforward generalization of the usual BCS-Gor'kov\ procedure \cite%
{PhysRevLett.3.552}. \ However, we include strong coupling in that we make
no assumptions about frequency cutoffs for the function $V\left( n,\vec{p}%
;n^{\prime },\vec{p}^{\prime }\right) $. \ This leads to a set of coupled
gap equations \
\begin{equation}
\Delta \left( n,\vec{p}\right) =-\sum_{n^{\prime }\vec{p}^{\prime }}F\left(
n^{\prime },\vec{p}^{\prime }\right) V\left( n,\vec{p};n^{\prime },\vec{p}%
^{\prime }\right) \Delta \left( n^{\prime },\vec{p}^{\prime }\right) ,
\label{eq:gap}
\end{equation}%
where $F(n,\vec{p})=\tanh [\beta E(n,\vec{p})/2]/\left[ 2E\left( n,\vec{p}%
\right) \right] $, $E(n,\vec{p})=[\xi ^{2}(n,\vec{p})+|\Delta (n,\vec{p}%
)|^{2}]^{1/2}$, and $\beta $ is the inverse temperature. \ Since the
experiments are done near the critical temperature, we linearize these
equations with respect to $\Delta $ and $F\left( n,\vec{p}\right) =\tanh %
\left[ \beta \xi \left( n,\vec{p}\right) /2\right] /\left[ 2\xi \left( n,%
\vec{p}\right) \right] $. \ In this case $F\left( n,\vec{p}\right) $ has the
full symmetry of the lattice. \

The point group of the system is $D_{6h}$ for UPt$_{3}$. The case of
interest is that of unconventional superconductivity. \ Let $R$ be a group
operation not the identity. \ We have that $\Delta \left( n,R\vec{p}\right)
\neq \Delta \left( n,\vec{p}\right) $ for all $n$. \ It is also true that $%
V\left( n,R\vec{p};n^{\prime },R\vec{p}^{\prime }\right) =V\left( n,\vec{p}%
;n^{\prime },\vec{p}^{\prime }\right) $ so $V$ can be decomposed into
channels corresponding to the irreducible representations of $G$. \ Regarded
as a function of $\vec{p},$ we seek the highest eigenvalue of $V,$ which
then determines the representation actually realized. Calculations for UPt$%
_{3}$ using experimental data to estimate $V\left( n,\vec{p};n^{\prime },%
\vec{p}^{\prime }\right) $ were done years ago, but were not conclusive \cite%
{PhysRevB.37.2372,PhysRevB.41.170} and first principles calculations using
the functional renormalization group have been done for other systems \cite%
{Wang17}, but not for UPt$_{3}$.

The split transition in UPt$_{3}$ \cite{Joynt88,PhysRevLett.62.1411} implies
that this representation is multi-dimensional, which for singlet
superconductivity means $E_{1g}$ or $E_{2g}$. \ We choose the former for
definiteness, but our conclusions apply equally to these two
representations. \ It is important to note that in the linear regime, Eq.~(%
\ref{eq:gap})\ determines the representation, but not which combination of
basis functions is chosen by the system. \ This degeneracy is broken at
higher order and there must be complex coefficients for a Kerr rotation to
occur. \ Thus we have that $\Delta\left( n,\vec{p}\right) =\Delta_{0}\left(
n,\vec{p}\right) ~p_{z}(p_{x}\pm ip_{y}),$ where $\Delta_{0}\left( n,R\vec{p}%
\right) =\Delta_{0}\left( n,\vec{p}\right) $ for all $R$. \

We may separate the bands into partially filled bands, of which there are 5
in UPt$_{3}$ indexed by $1\leq n\leq 5$ and completely filled or empty
bands, indexed by $n>5$. \ There are 6 separate Fermi surfaces in UPt$_{3}$.
\ For $n>5,$ $F\left( n,\vec{p}\right) $ is of order $1/B$ or less. \ We
expect $\left\vert \Delta \left( n^{\prime },\vec{p}\right) \right\vert $
for $n^{\prime }>5$ to be induced by a coupling $V\left( n,\vec{p};n^{\prime
},\vec{p}^{\prime }\right) $ that is off-diagonal in the band indices,
connecting partially filled to completely filled or completely empty bands.

Then there are two questions that are crucial for the calculation of $\theta
_{K}$. \ 1. How are the Ising-like variables $\pm $ in the equation $\Delta
\left( n,\vec{p}\right) =\Delta _{0}\left( n,\vec{p}\right) ~p_{z}(p_{x}\pm
ip_{y})$ determined as $n$ varies?\ \ 2. What is the order of magnitude of $%
\left\vert \Delta \left( n,\vec{p}\right) \right\vert $ for $n>5$?

1. The first question is fairly easy to answer in our model. $V\left( n,\vec{%
p};n^{\prime },\vec{p}^{\prime }\right) $ couples only bands with a $%
(p_{x}+ip_{y})$ with other bands with a $(p_{x}+ip_{y})$ gap and couples
only bands with a $(p_{x}-ip_{y})$ with other bands with a $(p_{x}-ip_{y})$
gap, \textit{i.e.}, it is diagonal in the $\pm $ degree of freedom. \
However, this coupling can be of either sign. \ Of course there are no
symmetries in the band index, so the couplings have no particular relation
to each other. \ In a Ginzburg-Landau approach, we may define $\Delta \left(
n,\vec{p}\right) =\Delta _{0}\left( n,\vec{p}\right) ~p_{z}(\eta
_{x}p_{x}+\eta _{y}p_{y})$ where the ``internal" order parameter $\vec{\eta}%
=\left( \eta _{x},\eta _{y}\right) $ depends on the band index. \ The free
energy in $E_{1g}$ is then
\begin{eqnarray}
F&=&\alpha _{m}~\vec{\eta}_{m}\cdot \vec{\eta}_{m}^{\ast }+\beta _{m}\left(
\vec{\eta}_{m}\cdot \vec{\eta}_{m}^{\ast }\right) ^{2}+\gamma _{m}\left\vert
\vec{\eta}_{m}\cdot \vec{\eta}_{m}\right\vert ^{2}  \notag \\
&+&J_{mn}\left( \vec{\eta}_{m}\cdot \vec{\eta}_{n}^{\ast }+\vec{\eta}%
_{m}^{\ast }\cdot \vec{\eta}_{n}\right) ,  \label{eq:GL}
\end{eqnarray}%
with a summation convention over the band indices $m$ and $n$ in effect. \
To break time-reversal symmetry we need the $\gamma _{m}$ to be positive,
and we need some of the $J_{mn}$ to be positive for some pair $\left(
m,n\right) $ of bands that differ in energy by 0.8 eV. \ Then we have a
problem of determining the ground state of an Ising magnet with more-or-less
random couplings. \ We may expect both $(p_{x}+ip_{y})~$and $(p_{x}-ip_{y})$
to occur in the absence of physical considerations to the contrary. \

2. The second question is more complicated. \ The size of $\left\vert \Delta
\left( n,\vec{p}\right) \right\vert $ for the partially occupied bands ($%
n\leq 5)$ is at least partially constrained by experiment. \ We expect at
least one and perhaps more of the gaps to be of order $2k_{B}T,$ \textit{i.e.%
}, about 10$^{-4}$eV. \ The superconductivity for $n>5$ is induced from the
partially occupied bands. \ For estimation purposes, we choose 2 bands from
Eq.~(\ref{eq:gap}), denoting them by $g$ for partially occupied and $e$ for
empty. \ We consider the separable forms: $V\left( g,\vec{p};g,\vec{p}%
^{\prime }\right) =-g_{g}f(\vec{p})f^{\ast }(\vec{p}^{\prime })$ and $%
V\left( g,\vec{p};e,\vec{p}^{\prime }\right) =-g_{e}f(\vec{p})f^{\ast }(\vec{%
p}^{\prime })$ with $f(\vec{p})=p_{z}\left( p_{x}+ip_{y}\right) $, so that $%
\Delta \left( g,\vec{p}\right) =\Delta _{g}^{\left( 0\right) }f(\vec{p})$
and $\Delta \left( e,\vec{p}\right) =\Delta _{e}^{\left( 0\right) }f(\vec{p}%
) $. The assumption that $\Delta _{e}^{\left( 0\right) }$ is induced means
that the corresponding component of $V\left( e,\vec{p};e,\vec{p}^{\prime
}\right) $ is small and we set it to zero. \ Then Eq.~(\ref{eq:gap}) yields%
\begin{eqnarray}
\Delta _{g}^{\left( 0\right) }\left[ 1-g_{g}F_{g}(\Delta _{g}^{\left(
0\right) }) \right] &=&\Delta _{e}^{\left( 0\right) }g_{e}~F_{e}(\Delta
_{e}^{\left( 0\right) })  \notag \\
\Delta _{e}^{\left( 0\right) } &=&\Delta _{g}^{\left( 0\right) }g_{e}\
F_{g}(\Delta _{g}^{\left( 0\right) }),  \label{eq:gapratio}
\end{eqnarray}%
where $F_{g}(\Delta _{g}^{\left( 0\right) }) =\sum_{\vec{p}}F(g,\vec{p})|f(%
\vec{p})|^{2}$ and $F_{e}(\Delta _{e}^{\left( 0\right) }) =\sum_{\vec{p}}F(e,%
\vec{p})|f(\vec{p})|^{2}$. At $T=0$ we have that $F_{g}(\Delta _{g}^{\left(
0\right)}) \approx N_{g}\left( 0\right) \ln \left( \omega _{c}/\Delta
_{g}^{\left( 0\right) }\right) $ while $F_{e}(\Delta _{e}^{\left( 0\right)
}) $ is of order $1/\omega _{c}$. The latter estimate also requires that the
cutoff $\omega _{c}$ is not too much less than the bandwidth, justified if
the interaction comes from the Coulomb interaction. \ We find that $%
\left\vert \Delta _{e}^{\left( 0\right) }/\Delta _{g}^{\left( 0\right)
}\right\vert \sim \left\vert g_{e}/g_{g}\right\vert$. \ Since all the bands
are $f$-like in UPt$_{3},$ the Coulomb matrix elements at short distances
are expected to be comparable, and this gives reason to suppose that $%
\left\vert \Delta _{e}^{\left( 0\right) }/\Delta _{g}^{\left( 0\right)
}\right\vert $ is of order unity.

In order to calculate $\theta_{K}$ we need the diagonal complex dielectric
function $\varepsilon_{xx}(\omega=0.8~\text{eV})$. \ This has been
determined by reflectivity measurements and a Kramers-Kronig analysis \cite%
{Schoenes85,PhysRevB.38.5338}. \ In this frequency range it is necessary to
include several Lorentz oscillators to fit the data, showing that there are
interband transitions at $\omega=0.8$ eV. \ This is in agreement with band
calculations\ \cite{Oguchi1985,PhysRevB.35.7260}. \ We extract the
approximate values $\mathrm{Re}\varepsilon_{xx}(\omega=0.8~\text{eV})
\approx3$ and $\mathrm{Im}\varepsilon_{xx}(\omega=0.8~\text{eV}) \approx25$
from these results.

The key quantity is of course the off-diagonal dielectric function $%
\varepsilon _{xy}\left( \omega =0.8~\text{eV}\right)$. \ The result for a
single pair of bands $(m,n)$ is
\begin{eqnarray}
&&\mathrm{Re}\varepsilon _{xy}\left( \omega \right) =\frac{64\pi }{\omega
^{2}}\sum_{\vec{p}}J_{x}^{\left( mn\right) }\left( \vec{p}\right)
J_{y}^{\left( mn\right) }\left( -\vec{p}\right)\times  \notag \\
&&\frac{\mathrm{Im}[\Delta \left( m,\vec{p}\right) \Delta ^{\ast }\left( n,%
\vec{p}\right)]}{E_{m}\left( \vec{p}\right) E_{n}\left( \vec{p}\right) }%
\delta \left[ \omega -E_{n}\left( \vec{p}\right) -E_{m}\left( \vec{p}\right) %
\right] ,  \label{eq:xy}
\end{eqnarray}%
obtained from the anomalous part of the lowest-order bubble diagram. \ This
is not the total dielectric function. To get that we must also sum over all
pairs. \ \ Here $J_{x,y}^{\left( mn\right) }\left( \vec{p}\right) $ are the
interband matrix elements of the current operator between single-particle
states in the partially occupied $m$-band and the empty $n$-band. \ An
analogous expression would hold for transitions from a completely full band
to a partially occupied band. \ This expression for $\varepsilon _{xy}$ is
to be compared to that for the familiar normal-state dielectric function%
\begin{eqnarray}
\mathrm{Im}\varepsilon _{xx}\left( \omega \right) &=&\frac{32\pi }{\omega
^{2}}\sum_{\vec{p}}J_{x}^{(mn)}\left( \vec{p}\right) J_{x}^{(mn)} \left(\vec{%
p}\right)\times  \notag \\
&& ~\delta \left[ \omega -\xi _{n}\left( \vec{p}\right) +\xi _{m}\left( \vec{%
p}\right) \right] .  \label{eq:xx}
\end{eqnarray}%
Although Eqs.~(\ref{eq:xy}) and (\ref{eq:xx}) contain quantities which are
poorly known, only the ratio is involved in the Kerr angle $\theta _{K}$. \
It is only this that allows us to give an order of magnitude estimate for $%
\theta _{K}$. \

To achieve this, we adopt a simple model of the bands in which the
single-particle energies $\xi _{m}\left( \vec{p}\right) $, $\xi _{n}\left(
\vec{p}\right) $ are random variables that are uniformly distributed over a
bandwidth $B,$ and the center of the $m$ and $n$ bands are separated by an
energy $B$. \ In the model the averages over the current matrix elements are
assumed to be the same for the two bands, and there are no correlations in
momentum space between the gap functions $\Delta \left( m,\vec{p}\right) $, $%
\Delta ^{\ast }\left( n,\vec{p}\right) $ and $J_{x}^{\left( mn\right)
}\left( \vec{p}\right) $, $J_{y}^{\left( mn\right) }\left( \vec{p}\right) $.
\ The computation of the ratio then reduces to a determination of the ratio
of the density of states parts of Eqs.~(\ref{eq:xy}) and (\ref{eq:xx}). \ \
The result is:%
\begin{eqnarray}
\frac{\mathrm{Re}\varepsilon _{xy}}{\mathrm{Im}\varepsilon _{xx}} &\sim &%
\frac{2\Delta _{m}\Delta _{n}}{\omega B}\ln \left( \frac{\omega _{c}}{\Delta
_{m}}\right) \times I_{s}  \notag \\
&\approx &5\times 10^{-7}I_{s}.  \label{eq:ratio}
\end{eqnarray}%
Here $\Delta _{m}$, $\Delta _{n}$ are the average values of $\left\vert
\Delta (m,\vec{p})\right\vert $, $\left\vert \Delta (n,\vec{p})\right\vert $%
, taken to be approximately equal to $2k_{B}T_{c},$ $\omega _{c}$ is the
cutoff for the gap $\Delta \left( m,\vec{p}\right) $ and we have assumed
that $\omega _{c}$ does not differ by orders of magnitude from $B$, which we
take to be $B=0.2$ eV. \ $I_{s}$ is the normalized angular integral over the
anisotropic gap functions. \ We discuss it further below. \ Eq.~(\ref%
{eq:ratio}) gives the contribution to the ratio $\mathrm{Re}\varepsilon
_{xy}/\mathrm{Im}\varepsilon _{xx}$ from one pair of bands. \ If we sum over
all bands and the ratio does not vary much from pair to pair, then we may
combine this value with the normal-state experimental value of $\varepsilon
_{xx}\left( \omega \right) $ quoted above to find $\theta _{K}\sim 2\times
10^{-7}$ at zero temperature, which is about 20\% or so of the value one
would get if the experimental results measured near $T_{c}$ are extrapolated
to $T=0$. Considering the approximations involved, and our general ignorance
about the mechanism of superconductivity, this is about all that can be
expected. \ Intraband theories typically give $\theta _{K}\sim (\Delta
/\omega )^{2}\sim 10^{-8},$ which is smaller.

These order-of-magnitude considerations all assume that $\theta _{K}$ does
not vanish by symmetry, which of course can happen if the angular integral
in Eq.~(\ref{eq:xy}) vanishes: $I_{s}=0$. \ Resolving this question amounts
to a symmetry analysis of the factor $J_{x}^{\left( mn\right) }\left( \vec{p}%
\right) J_{y}^{\left( mn\right) }\left( -\vec{p}\right) ~\left[ \Delta
\left( m,\vec{p}\right) \Delta ^{\ast }\left( n,\vec{p}\right) \right] $ for
each pair of bands that are separated by the laser photon energy 0.8 eV. \ \
Both the real and the imaginary part of $~\left[ \Delta \left( m,\vec{p}%
\right) \Delta ^{\ast }\left( n,\vec{p}\right) \right] $ contribute to $%
\theta _{K}$. \ The product $J_{x}J_{y}$ transforms according to the $E_{2g}$
representation of $D_{6h}$. \ The representation of $\left[ \Delta \left( m,%
\vec{p}\right) \Delta ^{\ast }\left( n,\vec{p}\right) \right] $ is in
general reducible, and if we define it as $\Gamma _{\Delta },$ then the
integral vanishes if and only if the $E_{2g}\times \Gamma _{\Delta }$
representation does not contain the $A_{1g}$ (identity) representation. \ $%
\theta _{K}$ itself will vanish if and only if the integral in Eq.~(\ref%
{eq:xy}) vanishes for every pair of bands. \ Again we will use the example
that $\Delta \left( m,\vec{p}\right) $ and $\Delta \left( n,\vec{p}\right) $
both transform as the $E_{1g}$ representation, but the considerations apply
to all representations. \ For this case we have essentially two
possibilities. $\ $The first is $\Delta \left( m,\vec{p}\right) \Delta
^{\ast }\left( n,\vec{p}\right) \sim p_{z}^{2}\left( p_{x}+ip_{y}\right)
\left( p_{x}+ip_{y}\right) ^{\ast }=p_{z}^{2}(p_{x}^{2}+p_{y}^{2}),$ and we
find $\Gamma _{\Delta }=A_{1g}$ and since $E_{2g}\times A_{1g}=E_{2g},$ $%
I_{s}=0$ and there is no contribution to $\theta _{K}$ for a pair of bands
both of which are of the $\left( p_{x}+ip_{y}\right) $ type. \ The second
case is $\Delta \left( m,\vec{p}\right) \Delta ^{\ast }\left( n,\vec{p}%
\right) \sim p_{z}^{2}\left( p_{x}+ip_{y}\right) \left( p_{x}-ip_{y}\right)
^{\ast }=p_{z}^{2}\left( p_{x}^{2}-p_{y}^{2}+2ip_{x}p_{y}\right) ,$ which
gives $\Gamma _{\Delta }=E_{2g}$. \ Since $E_{2g}\times
E_{2g}=A_{1g}+A_{2g}+E_{2g}$, $I_{s}\neq 0$ a pair of this type can
contribute to nonzero $\theta _{K}$. \ Thus, continuing the analogy with the
Ising model, each band being of the $\pm $ type, we see that a
\textquotedblleft ferromagnetic" state with all bands of the $+$ type or all
bands being of the $-$ type does not lead to a Kerr rotation. \ We must have
a mixture of $"+"$ and $"-"$ gaps on different bands. \ Qualitatively, we
may perhaps think of the Kerr rotation as the creation of a broken pair that
must be able to absorb both the energy and the angular momentum of the
photon - this can only occur if the chirality of the two gaps involved is
different.

The observation of a nonzero Kerr rotation in UPt$_{3}$ is seen to have a
multitude of consequences, going well beyond the fact that it demands that
there must be time-reversal symmetry breaking. \ There must also exist a
pairing field in a completely filled or a completely empty band, since the
photon energy exceeds the bandwidth. \ The observation of this pairing field
is novel: in fact we believe it is the first time that it has been observed
in experiments in a band that is very distant in energy from the Fermi
energy. \ The existence of this pairing field is also not enough to explain
the phenomenon: there must be a subtle pattern of relative symmetry between
the bands involved. \ Finally, we note that if there is superconductivity
with opposite chiralities $p_{z}\left( p_{x}+ip_{y}\right) $ and $%
p_{z}\left( p_{x}-ip_{y}\right) $ coexisting in the same sample, it raises
interesting questions concerning topology issues. \ How does the coexistence
modify the theory of Majorana fermions in the vortex cores, and protected
surface states?

We thank Po-Ho Yang and A. V. Chubukov for useful discussions. Financial
support from MOST of Taiwan (Grant No. 105-2112-M-003-005) is acknowledged.


%

\end{document}